\documentclass[10pt]{blois07}
\usepackage{graphicx}
\usepackage{cite,./mcite}

\begin{document}
\title{Gluon Saturation and Black Hole Criticality}
\author{L.~{\'A}lvarez-Gaum{\' e}$^1$, 
C.~G{\' o}mez$^2$\protect\footnote{~~talk presented at EDS07}, 
A.~Sabio~Vera$^1$, A.~Tavanfar$^{1,3}$, M.~A.~V{\'a}zquez-Mozo$^4$}
\institute{$^1$Theory Group, Physics Department, CERN, CH-1211 Geneva 23, 
Switzerland,\\ 
$^2$Instituto de F{\' \i}sica Te{\' o}rica UAM/CSIC, Universidad Aut{\' o}noma 
de Madrid, E-28049 Madrid, Spain,\\
$^3$Institute for Studies in Theoretical Physics and Mathematics (IPM) P.O. 
Box 19395-5531, Tehran, Iran,\\
$^4$Departamento de F{\' \i}sica Fundamental, Universidad de Salamanca, Plaza 
de la Merced s/n, E-37008 Salamanca, Spain and Instituto Universitario de 
F{\' \i}sica Fundamental y Matem{\'a}ticas (IUFFyM), Universidad de Salamanca, 
Spain.}
\maketitle
\begin{abstract}
We discuss the recent proposal~\cite{AlvarezGaume:2006dw} where it was 
shown that the critical anomalous dimension associated to the onset of 
non--linear effects in the high energy limit of QCD coincides with the 
critical exponent governing the radius of the black hole formed in the 
spherically symmetric collapse of a massless scalar field. We argue that a 
new essential ingredient in this mapping between gauge theory and gravity is 
continuous self-similarity, not present in the scalar field case but in the 
spherical collapse of a perfect fluid with barotropic equation of state. We 
identify this property with geometric scaling, present in DIS data at small 
values of Bjorken $x$. We also show that the Choptuik exponent in dimension five 
tends to the QCD critical value in the traceless limit of the energy momentum tensor.
\end{abstract}

\section{Criticality in high energy QCD and black hole formation}
\label{sec:Introduction}

One of the major insights of string theory is the unexpected connection 
between black hole physics and confinement in QCD. This connection is 
realized on the basis of the deep 
holographic~\cite{tHooft:1993gx,*Susskind:1994vu} duality 
between gravity and gauge 
theories~\cite{Maldacena:1997re,*Gubser:1998bc,*Aharony:1999ti,Witten:1998qj}.
 A particularly interesting connection between black holes and 
gauge theories is the dual interpretation~\cite{Witten:1998qj} 
of the Hawking--Page phase transition~\cite{Hawking:1982dh} in gravity as 
the confinement / deconfinement transition in gauge theory at finite 
temperature. 
\begin{figure}[h]
\centerline{\includegraphics[width=4cm,angle=0]{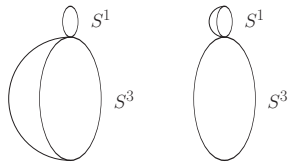}}
\caption{The two relevant bulk geometries for confinement / deconfinement 
transition.}
\label{geometries}
\end{figure} 
Holography is based on a very concrete set of rules for computing quantum 
field theory expectation values in terms of classical solutions to SUGRA 
equations with concrete boundary conditions. A further important ingredient 
of holography is the generalization of the gauge--gravity correspondence to 
non local observables such as Wilson or Polyakov loops. In this `non local'  
version of the correspondence expectation values of Wilson loops are defined 
as sums of string world sheets in the bulk geometry with boundary determined 
by the loop. 

\begin{figure}[h]
\begin{center}
  \includegraphics [bb= 23 247 572 666,width=8cm]      
  {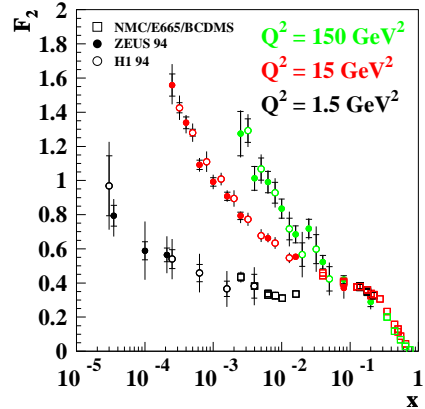}
\end{center}
\caption {The dependence on $x$ of the proton structure function $F_2$ for 
different values of $Q{}^{2}$.}
\label{HERA}
\end{figure}

\begin{figure}[h]
\begin{center}
\hspace{-2cm}  \includegraphics [width=8cm]      
  {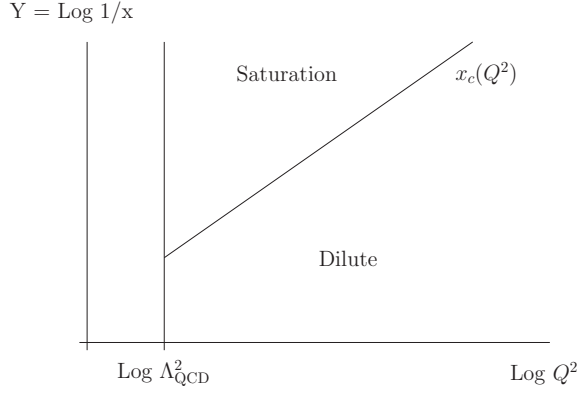}
\end{center}
\caption {QCD kinematic space in DIS}
\label{kinematic}
\end{figure}

Wilson loops are natural candidates to define the order parameter 
of QCD phases. This is the case not only for confinement but also for 
the transition from a dilute gas of partons to the so--called `color glass 
condensate'~\cite{Iancu:2000hn,*Iancu:2001ad} of hadronic parton 
distributions. 
In the same way as there exists a geometrical qualitative picture of the 
transition to confinement, illustrated in Fig.~\ref{geometries}, it is natural 
to search for the corresponding holographic description of the saturation 
phenomena present in the high energy limit of QCD~\cite{Stasto:2000er}. A very 
important experimental discovery in HERA data was the rise of the gluon 
distribution function $x \,G (x,Q{}^{2})$ (which is related to the number of 
gluons in the target proton wave function with effective transverse size of 
order $1/Q{}^{2}$ carrying a fraction $x$ of the hadron longitudinal momentum) 
when, for fixed $Q{}^{2}$, the value of $x$ becomes small. This can be 
seen, {\it e.g.}, in Fig.~\ref{HERA}.
In the `dipole frame' the total cross section for the scattering of the 
virtual photon off the hadron can be expressed in terms of the probability 
amplitude for the photon to decay into a quark--antiquark pair, creating a 
colour dipole of size $r = 1/Q$, which then scatters off the proton's 
effective colour field. The forward scattering amplitude for the dipole 
depends on $r$ and the rapidity variable $Y = \log (1/x)$. In the leading 
order approximation this scattering amplitude depends linearly on 
$x \,G (x,Q{}^{2})$. Since unitarity requires the forward scattering amplitude 
not to be larger than one this indicates that the rise of the gluon 
distribution should reach a saturation point, leading to the kinematic diagram 
shown in Fig.~\ref{kinematic}, where the saturation line indicates the 
critical value $x_c (Q{}^{2})$ such that for $x < x_c$ and fixed $Q{}^{2}$ the gluon 
distribution function becomes effectively constant in $x$. 

\begin{figure}[h]
\begin{center}
\hspace{-2cm}  \includegraphics [width=5cm]      
  {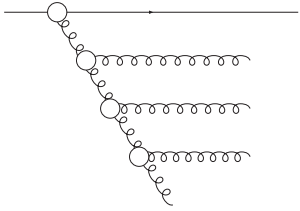}
\end{center}
\caption {BFKL gluon cascade in multi--Regge kinematics}
\label{cascade}
\end{figure}

The two main theoretical problems associated with the previous picture are to 
identify $i)$ the dynamical origin of the rise of $x \,G (x,Q{}^{2})$ with 
decreasing $x$ and $ii)$ the nature of the non--linear dynamics responsible 
for saturation and restoration of unitarity. The current understanding is that 
the dynamical origin of the rise of the gluon distribution 
function is due to the dominance of BFKL 
dynamics~\cite{Fadin:1975cb,*Kuraev:1976ge,*Kuraev:1977fs,*Balitsky:1978ic} 
while the non--linear BK equation~\cite{Balitsky:1995ub,*Kovchegov:1999yj} is 
responsible for the onset of saturation effects in the high energy limit of 
scattering amplitudes. In the eikonal approximation the exponential rise in 
$Y$ for the forward scattering amplitude can be computed as the Wilson loop 
for the quark--antiquark pair propagating in the effective colour field of the 
proton which, in the proton infinite momentum frame and large center--of--mass 
energies, is dominated by soft gluon emissions in multi--Regge kinematics, 
with strong ordering in longitudinal components but not in transverse ones. 
These configurations, shown in Fig.~\ref{cascade}, build up 
the BFKL hard pomeron.

\begin{figure}[h]
\begin{center}
\hspace{-2cm}  \includegraphics [width=9cm] {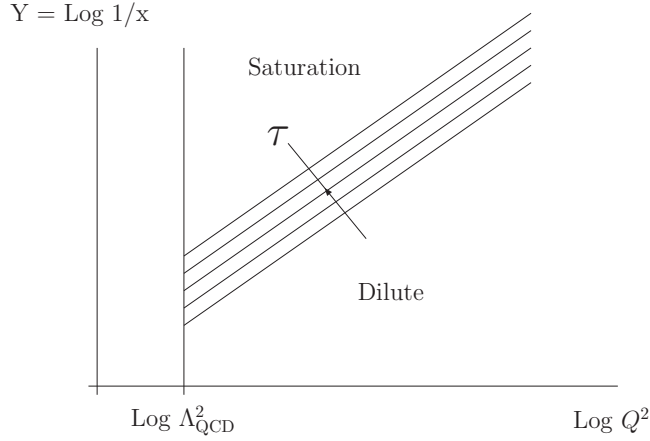}
\end{center}
\caption {Continuous self--similarity on the $(Y, \log Q{}^{2})$ plane.}
\label{Saturation}
\end{figure}

\begin{figure}[h]
\begin{center}
\hspace{1cm}\includegraphics [width=8cm]      
  {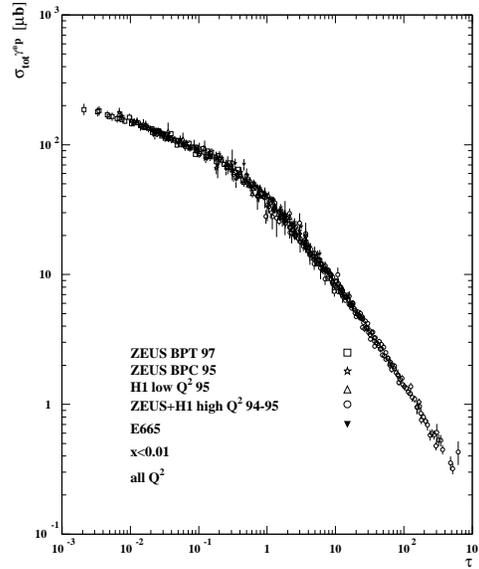}
\end{center}
\vspace{-0.5cm}
\caption{HERA data for $\sigma_{\gamma^* p}$ with $x<0.01$ versus the variable 
$\tau$.}
\label{HERAtau}
\end{figure}

In this framework the rapidity $Y$ acts as a cutoff in the effective 
integration over longitudinal momenta and the BFKL equation 
controlling the evolution in $Y$ plays the conceptual r{\^o}le of a 
renormalization group equation with a `fixed point', generated by non--linear 
effects, at the saturation line. This line, of the form 
$Q_s^2 (x) \sim x^{-\lambda}$, is characterized by the `saturation exponent', 
which, in the limit of a very small coupling, reads $\lambda 
\simeq \alpha_s N_c 2.44 / \pi$. A direct consequence of the onset of 
non--linear effects is that asymptotic amplitudes only depend on the variable 
$\tau \simeq Q{}^{2} x^\lambda$ nearby the saturation region, {\it i.e.}, in the 
$(Y, \log Q{}^{2})$ plane this implies that physical observables only depend 
on lines of constant $\tau$ as it is shown in Fig.~\ref{Saturation}. Along 
those lines any continuous boost in longitudinal components can be compensated 
by an equivalent one in the transverse directions to leave physical 
quantities invariant. 
The `geometric scaling' on this variable has been experimentally observed 
in HERA data~\cite{Stasto:2000er} for the $\gamma^* p$ cross sections in 
the region $x < 0.01$ over a large range in $Q{}^{2}$, see Fig.\ref{HERAtau}.

Our target is to find the holographic dual of the saturation 
line~\cite{Nosotros}. Taking into 
account the Wilson loop picture of the forward scattering amplitude 
for the dipole, the holographic representation of this quantity leads us to 
sum over world sheet amplitudes for a certain bulk 
geometry~\cite{Liu:2006ug,*Liu:2006nn,*Liu:2006he}. Since the 
dynamics we want to describe in gravity dual terms is the rapidity dependence 
of the amplitude, we will formally consider a background metric depending on 
the dual variable to $Y$ in such a way that in the transition 
from the dilute to the dense or saturated regime it manifests some sort 
of `geometric scaling' in terms of properly chosen holographic variables. 
To establish the correspondence correctly this scaling must be characterized 
by a critical quantity related to the saturation exponent $\lambda$.

A first hint in this direction was shown in~\cite{AlvarezGaume:2006dw}. There 
it was argued that in the numerical studies of black hole formation for the 
spherically symmetric collapse of a massless scalar field carried out by 
Choptuik~\cite{Choptuik:1992jv} (see~\cite{Gundlach:2002sx} for a review) 
there 
appears a critical exponent very similar to $\lambda$. In particular, if we 
denote by $p$ a generic parameter describing the initial radial density for 
imploding scalar waves, Choptuik found that there are critical lines in $p$, 
$p = p^*$, such that, if $p<p^*$ the scalar wave packet implodes through 
$r=0$ and then disperses into flat space--time. But if $p>p*$, {\it i.e.}, 
the `supercritical' case, then after the implosion there is a fraction of 
field which forms a small black hole. The interesting point is that its radius 
scales as $r_{\rm BH} \simeq |p-p^*|^{\frac{1}{\lambda_c}}$, and it turns out 
that precisely in dimension five $\lambda_c \simeq 2.44$. 

\begin{figure}[h]
\hspace{-1.5cm}\includegraphics [width=6cm,angle=-90]{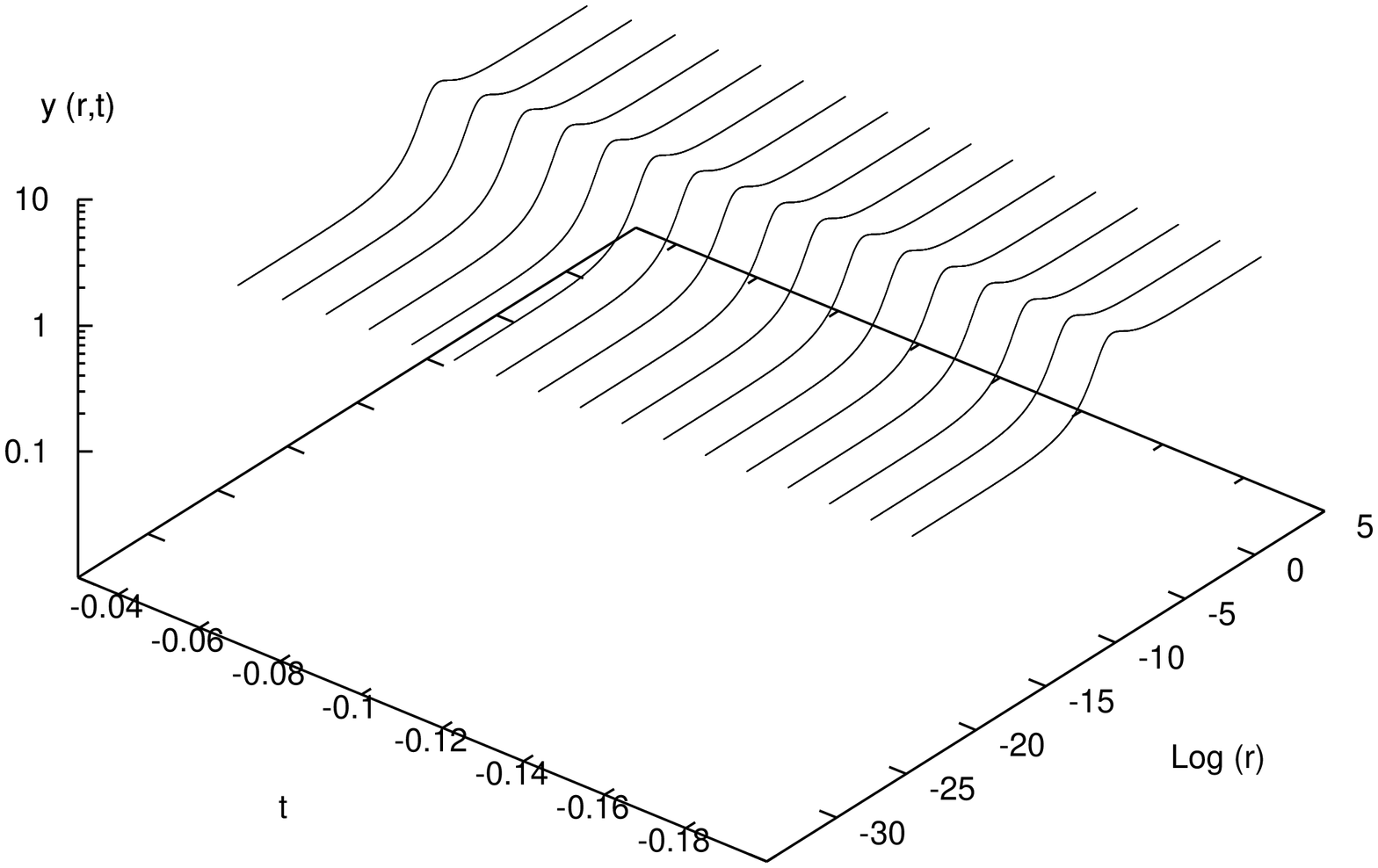}\includegraphics [width=6cm,angle=-90] {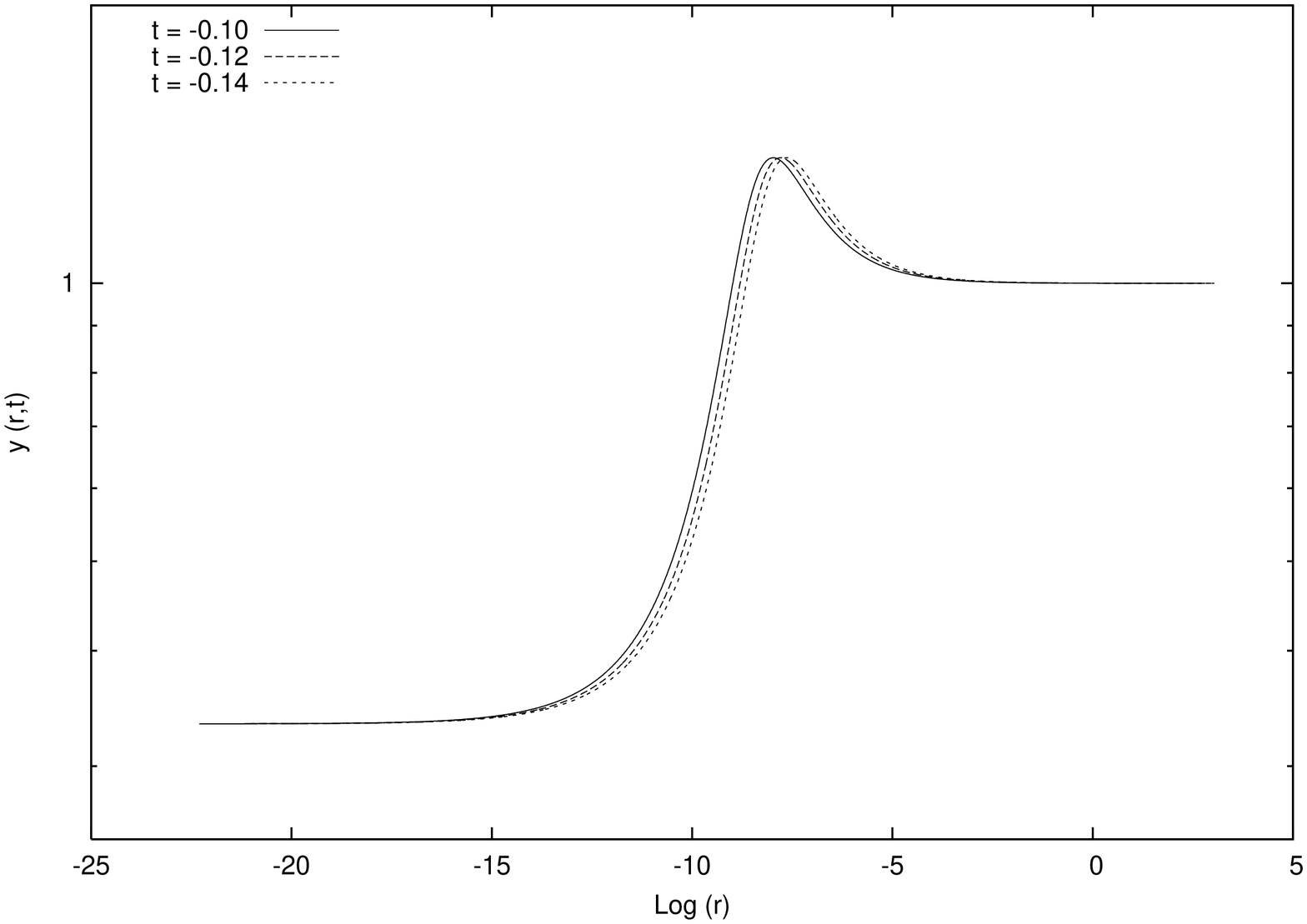}
\caption{A typical solution to gravitational collapse of a perfect fluid with 
continuous self--similarity.}
\label{CSSyrt}
\end{figure}

\begin{figure}[h]
\hspace{-1.5cm}\includegraphics [width=6cm,angle=-90]{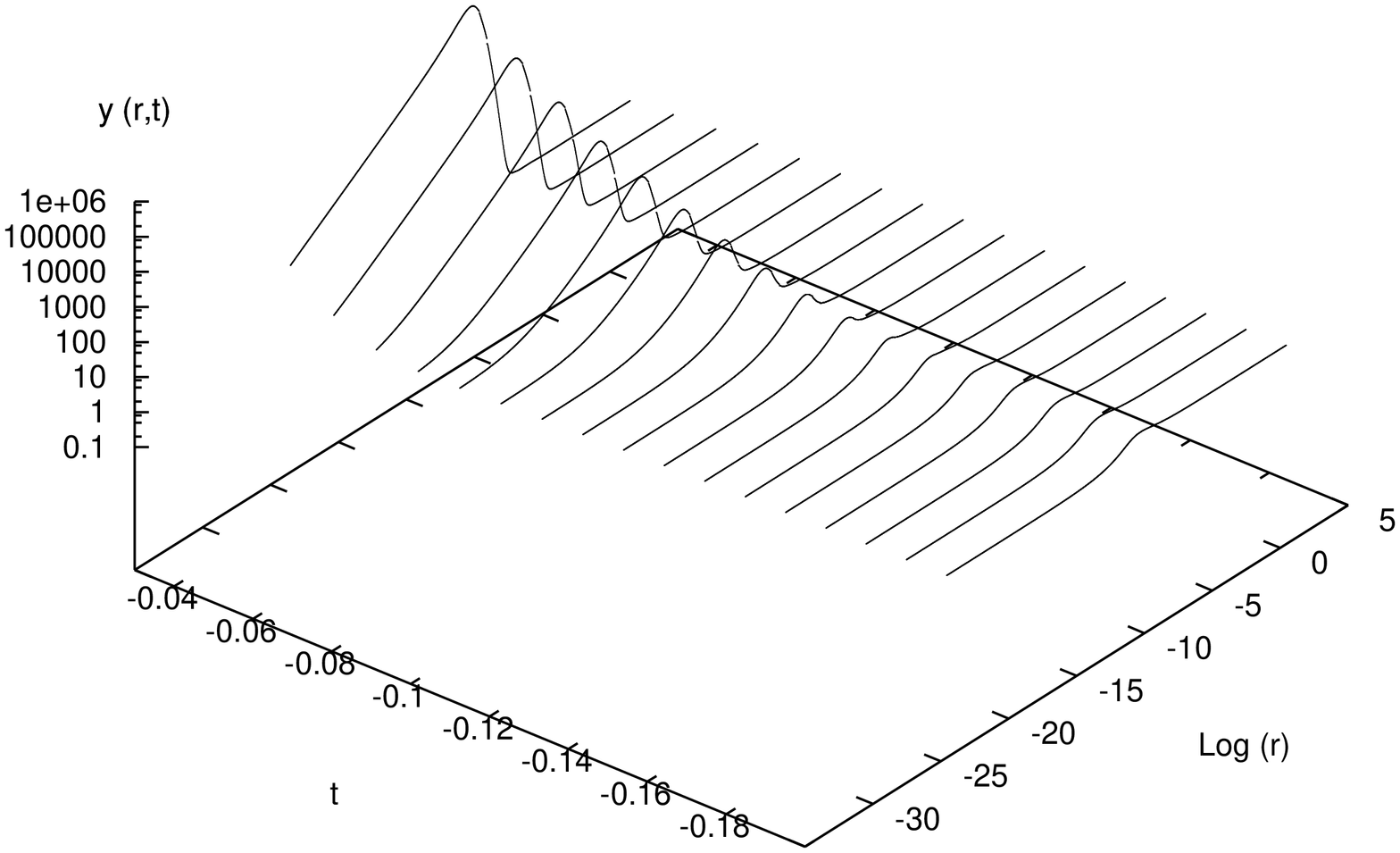}\includegraphics [width=6cm,angle=-90] {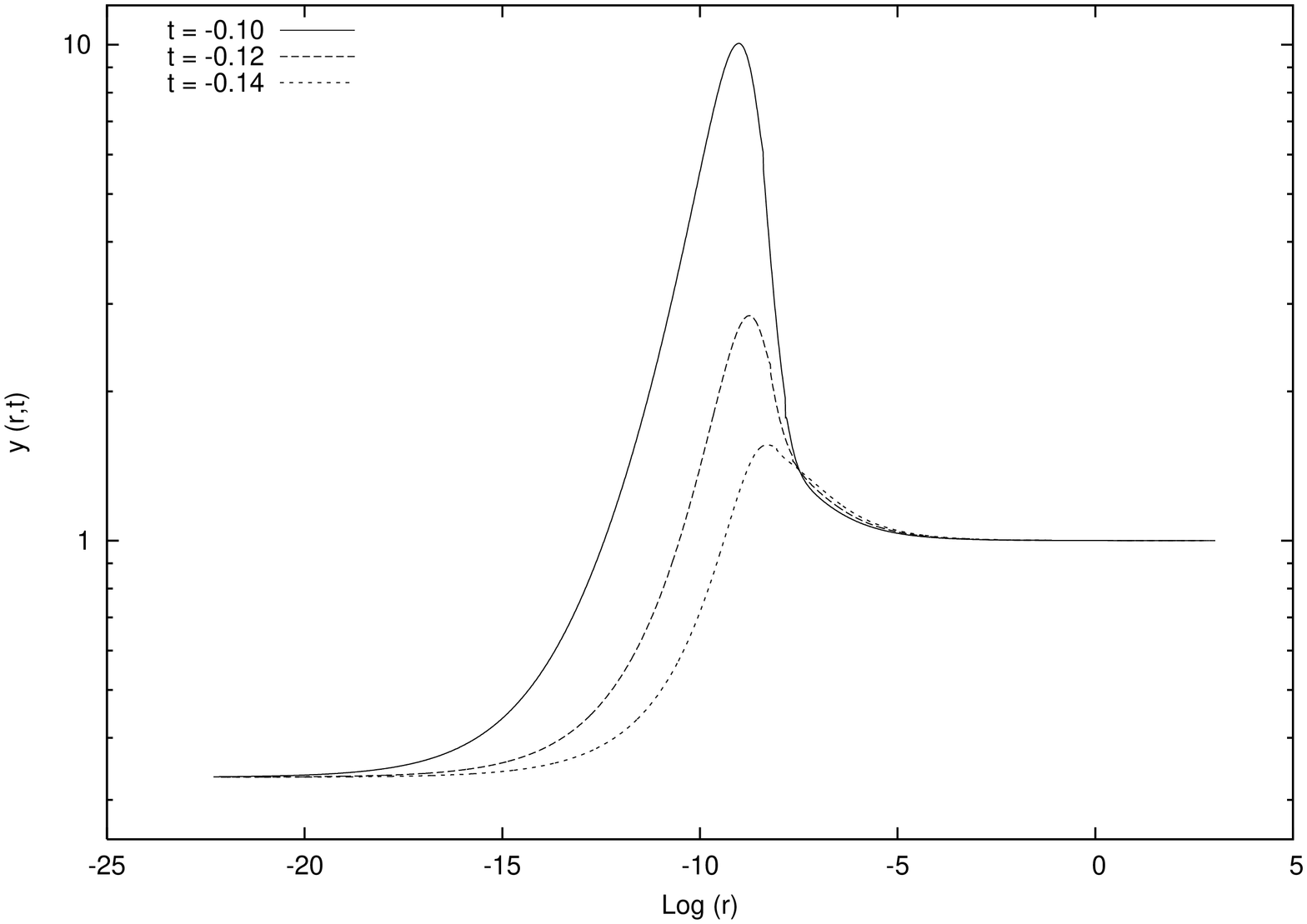}
\caption{Continuous self--similarity is lost by an exponentially growing mode 
in $t$.}
\label{pertCSSyr}
\end{figure}

Nonetheless, there is a difficulty to map the collapse of a scalar field with 
QCD, namely, the metric and field components obtained in this case manifest 
`discrete self--similarity'. This means that a similar variable as the 
above--mentioned $\tau$ leaves physical observables invariant under the 
transformation $\tau \rightarrow \tau + \Delta$, with $\Delta$ a constant 
which has no analogue in four dimensional high energy scattering. We have 
investigated in detail a different type of gravitational collapse which has 
self--similarity, in this case `continuous self--similarity' (CSS): the 
spherical collapse of a perfect fluid with a barotropic equation of state. The 
Einstein's equations together with matter's equations of motion in $d$ 
dimensions can be solved assuming a unique dependence on the variable 
$\tau=-r/t$. As an example we discuss the function $y(r,t)$ which is 
proportional to the ratio of the mean density inside the sphere of radius $r$ 
to the local density at $r$. In Fig.~\ref{CSSyrt} it can be seen how $y(r,t)$ 
maintains a constant $r$--profile for different values of $t$. This implies 
that 
the solution is CSS since any change in the time coordinate can be compensated 
by a change in $r$ keeping $y$ unchanged. This CSS property is what we 
associate with `geometric scaling' in QCD, with $t$ and $r$ being, respectively, 
the holographic duals of $\alpha_s N_c Y $ and $\log{Q{}^{2}}$ in QCD.
The Choptuik exponent characterizing the black hole radius can be obtained by 
searching for Liapunov modes of instability of the CSS solution. In 
Fig.~\ref{pertCSSyr} we show how the exponentially growing mode removes 
CSS from the solution to the collapse. The rate of growth of this mode is 
given by a coefficient which coincides with Choptuik's exponent. We have 
numerically extracted this coefficient in the five dimensional case and proven 
that it is very close to the QCD saturation exponent in the limit of traceless 
energy--momentum tensor for the fluid~\cite{Nosotros}.

\begin{footnotesize}
\bibliographystyle{blois07} 
{\raggedright
\bibliography{blois07}

\providecommand{\etal}{et al.\xspace}
\providecommand{\href}[2]{#2}
\providecommand{\coll}{Coll.}
\catcode`\@=11
\def\@bibitem#1{%
\ifmc@bstsupport
  \mc@iftail{#1}%
    {;\newline\ignorespaces}%
    {\ifmc@first\else.\fi\orig@bibitem{#1}}
  \mc@firstfalse
\else
  \mc@iftail{#1}%
    {\ignorespaces}%
    {\orig@bibitem{#1}}%
\fi}%
\catcode`\@=12
\begin{mcbibliography}{10}

\bibitem{AlvarezGaume:2006dw}
L.~{\'A}lvarez-Gaum{\'e}, C.~G{\'o}mez, and M.~A. V{\'a}zquez-Mozo,
\newblock Phys. Lett.{} {\bf B649},~478~(2007).
\newblock \href{http://www.arXiv.org/abs/hep-th/0611312}{{\tt
  hep-th/0611312}}\relax
\relax
\bibitem{tHooft:1993gx}
G.~'t~Hooft~(1993).
\newblock \href{http://www.arXiv.org/abs/gr-qc/9310026}{{\tt
  gr-qc/9310026}}\relax
\relax
\bibitem{Susskind:1994vu}
L.~Susskind,
\newblock J. Math. Phys.{} {\bf 36},~6377~(1995).
\newblock \href{http://www.arXiv.org/abs/hep-th/9409089}{{\tt
  hep-th/9409089}}\relax
\relax
\bibitem{Maldacena:1997re}
J.~M. Maldacena,
\newblock Adv. Theor. Math. Phys.{} {\bf 2},~231~(1998).
\newblock \href{http://www.arXiv.org/abs/hep-th/9711200}{{\tt
  hep-th/9711200}}\relax
\relax
\bibitem{Gubser:1998bc}
S.~S. Gubser, I.~R. Klebanov, and A.~M. Polyakov,
\newblock Phys. Lett.{} {\bf B428},~105~(1998).
\newblock \href{http://www.arXiv.org/abs/hep-th/9802109}{{\tt
  hep-th/9802109}}\relax
\relax
\bibitem{Aharony:1999ti}
O.~Aharony, S.~S. Gubser, J.~M. Maldacena, H.~Ooguri, and Y.~Oz,
\newblock Phys. Rept.{} {\bf 323},~183~(2000).
\newblock \href{http://www.arXiv.org/abs/hep-th/9905111}{{\tt
  hep-th/9905111}}\relax
\relax
\bibitem{Witten:1998qj}
E.~Witten,
\newblock Adv. Theor. Math. Phys.{} {\bf 2},~253~(1998).
\newblock \href{http://www.arXiv.org/abs/hep-th/9802150}{{\tt
  hep-th/9802150}}\relax
\relax
\bibitem{Hawking:1982dh}
S.~W. Hawking and D.~N. Page,
\newblock Commun. Math. Phys.{} {\bf 87},~577~(1983)\relax
\relax
\bibitem{Iancu:2000hn}
E.~Iancu, A.~Leonidov, and L.~D. McLerran,
\newblock Nucl. Phys.{} {\bf A692},~583~(2001).
\newblock \href{http://www.arXiv.org/abs/hep-ph/0011241}{{\tt
  hep-ph/0011241}}\relax
\relax
\bibitem{Iancu:2001ad}
E.~Iancu, A.~Leonidov, and L.~D. McLerran,
\newblock Phys. Lett.{} {\bf B510},~133~(2001).
\newblock \href{http://www.arXiv.org/abs/hep-ph/0102009}{{\tt
  hep-ph/0102009}}\relax
\relax
\bibitem{Stasto:2000er}
A.~M. Stasto, K.~J. Golec-Biernat, and J.~Kwiecinski,
\newblock Phys. Rev. Lett.{} {\bf 86},~596~(2001).
\newblock \href{http://www.arXiv.org/abs/hep-ph/0007192}{{\tt
  hep-ph/0007192}}\relax
\relax
\bibitem{Fadin:1975cb}
V.~S. Fadin, E.~A. Kuraev, and L.~N. Lipatov,
\newblock Phys. Lett.{} {\bf B60},~50~(1975)\relax
\relax
\bibitem{Kuraev:1976ge}
E.~A. Kuraev, L.~N. Lipatov, and V.~S. Fadin,
\newblock Sov. Phys. JETP{} {\bf 44},~443~(1976)\relax
\relax
\bibitem{Kuraev:1977fs}
E.~A. Kuraev, L.~N. Lipatov, and V.~S. Fadin,
\newblock Sov. Phys. JETP{} {\bf 45},~199~(1977)\relax
\relax
\bibitem{Balitsky:1978ic}
I.~I. Balitsky and L.~N. Lipatov,
\newblock Sov. J. Nucl. Phys.{} {\bf 28},~822~(1978)\relax
\relax
\bibitem{Balitsky:1995ub}
I.~Balitsky,
\newblock Nucl. Phys.{} {\bf B463},~99~(1996).
\newblock \href{http://www.arXiv.org/abs/hep-ph/9509348}{{\tt
  hep-ph/9509348}}\relax
\relax
\bibitem{Kovchegov:1999yj}
Y.~V. Kovchegov,
\newblock Phys. Rev.{} {\bf D60},~034008~(1999).
\newblock \href{http://www.arXiv.org/abs/hep-ph/9901281}{{\tt
  hep-ph/9901281}}\relax
\relax
\bibitem{Nosotros}
L.~{\'A}lvarez-Gaum{\'e}, C.~G{\'o}mez, A.~Sabio~Vera, A.~Tavanfar, and M.~A.
  V{\'a}zquez-Mozo,
\newblock work in progress{}\relax
\relax
\bibitem{Liu:2006ug}
H.~Liu, K.~Rajagopal, and U.~A. Wiedemann,
\newblock Phys. Rev. Lett.{} {\bf 97},~182301~(2006).
\newblock \href{http://www.arXiv.org/abs/hep-ph/0605178}{{\tt
  hep-ph/0605178}}\relax
\relax
\bibitem{Liu:2006nn}
H.~Liu, K.~Rajagopal, and U.~A. Wiedemann,
\newblock Phys. Rev. Lett.{} {\bf 98},~182301~(2007).
\newblock \href{http://www.arXiv.org/abs/hep-ph/0607062}{{\tt
  hep-ph/0607062}}\relax
\relax
\bibitem{Liu:2006he}
H.~Liu, K.~Rajagopal, and U.~A. Wiedemann,
\newblock JHEP{} {\bf 03},~066~(2007).
\newblock \href{http://www.arXiv.org/abs/hep-ph/0612168}{{\tt
  hep-ph/0612168}}\relax
\relax
\bibitem{Choptuik:1992jv}
M.~W. Choptuik,
\newblock Phys. Rev. Lett.{} {\bf 70},~9~(1993)\relax
\relax
\bibitem{Gundlach:2002sx}
C.~Gundlach,
\newblock Phys. Rept.{} {\bf 376},~339~(2003).
\newblock \href{http://www.arXiv.org/abs/gr-qc/0210101}{{\tt
  gr-qc/0210101}}\relax
\relax
\end{mcbibliography}
}
\end{footnotesize}
\end{document}